\documentstyle[11pt,newpasp,twoside]{article}
\input{psfig}
\def\HI{H{\,\small I}}
\def\HII{H{\,\small II}}
\newcommand{\ltsima} {$\; \buildrel < \over \sim \;$}
\newcommand{\gtsima} {$\; \buildrel > \over \sim \;$}
\newcommand{\lta} {\lower.5ex\hbox{\ltsima}}
\newcommand{\gta} {\lower.5ex\hbox{\gtsima}}

\def\edcomment#1{\iffalse\marginpar{\raggedright\sl#1\/}\else\relax\fi}
\reversemarginpar

\begin{document}
\title{Tidal Remnants and Intergalactic \HII\ Regions}
 \author{Tom Oosterloo, Raffaella Morganti}
\affil{Netherlands Foundation for Research in Astronomy, Dwingeloo, NL}
\author{Elaine M. Sadler}
\affil{School of Physics, University of Sydney, NSW 2006, Australia}
\author{Annette  Ferguson}
\affil{Max-Planck Institute f\"ur Astrophysik, Karl-Schwarzschild-Str. 1,
85748 Garching, Germany}
\author{Thijs van der Hulst}
\affil{Kapteyn Astronomical Institute, RuG, Landleven 12, 9747 AD,
Groningen, NL}
\author{Helmut Jerjen}
\affil{Research School of Astronomy \& Astrophysics, Australian National
University, ACT 2611, Canberra, Australia}

\begin{abstract}

We report the discovery of two small intergalactic \HII\ regions in
the loose group of galaxies around the field elliptical NGC 1490. The
\HII\ regions are located at least 100 kpc from any optical galaxy but
are associated with a number of large \HI\ clouds that are lying along an arc
500 kpc in length and that have no optical counterpart on the Digital Sky
Survey.  The sum of the \HI\ masses of the clouds is almost 10$^{10}$
M$_\odot$ and the largest \HI\ cloud is about 100 kpc in size.  Deep optical
imaging reveals a very low surface brightness counterpart to this largest \HI\
cloud, making this one of the \HI\ richest optical galaxies known ($M_{\rm
HI}/L_{\rm V} \sim 200$).  Spectroscopy of the \HII\ regions indicates that
the abundance in these \HII\ regions is only slightly sub-solar, excluding a
primordial origin of the \HI\ clouds.  The \HI\ clouds are perhaps remnants
resulting from the tidal disruption of a reasonably sized galaxy, probably
quite some time ago, by the loose group to which NGC~1490 belongs.
Alternatively, they are remnants of the merger that created the field
elliptical NGC~1490.  The isolated \HII\ regions show that star formation on a
very small scale can occur in intergalactic space in gas drawn from galaxies
by tidal interactions.  Many such intergalactic small star formation regions
may exist near tidally interacting galaxies.

\end{abstract}

\section{Introduction}

Hierarchical accretion and merging of small clumps appears to be a good
description of the formation of early-type galaxies.  In many early-type
galaxies signatures of recent hierarchical assembly can be observed,
particularly in field galaxies.  Optical fine structure -shells, dust lanes,
``X'' structures- revealed through un-sharp masking of optical images suggests
a recent accretion event.  These fine structures are correlated with the
presence of an intermediate age (1-2 Gyr) population of stars (e.g.\ Schweizer
et al.\ 1990).  Moreover, comparing detailed stellar population models with
high signal-to-noise optical spectra indicates the presence of an intermediate
age population in several early-type galaxies (e.g.\ Trager et al.\ 2000).  A
small but significant fraction of early-type galaxies contain neutral
hydrogen.  The origin of this neutral hydrogen is generally thought to be
external and due to recent accretion of companions.  Hence the presence of
\HI\ in early-type galaxies is generally thought to fit into the picture of
continued accretion of companions and the associated intermittent episodes of
star formation in some early-type galaxies.

We are undertaking a systematic \HI\ survey, using the Australia Telescope
Compact Array (ATCA), of {\sl all} early-type galaxies south of $\delta <
-25^\circ$ with$ V<6000$ km s$^{-1}$ (based on HIPASS, the Parkes All Sky \HI\
Survey, see Barnes et al. 2001 and elsewhere in this volume) to obtain an
unbiased sample of \HI-rich early-type galaxies. Such a complete sample will
allow us to study the effects of \HI\ on the evolution of early-type galaxies
in different environments, and the relation between \HI\ and other indicators
of recent accretion and star formation (see also Sadler et al. 2002). We can
also study whether \HI\ always has an external origin in early-type galaxies
or whether in some early-type galaxies the \HI\ may be related to an earlier
phase in the evolution of the galaxies.

During this HIPASS follow-up program we discovered two intergalactic
\HII\ regions  near the field elliptical NGC 1490 that are probably
related to merging/accretion within the loose group of galaxies around
NGC~1490, or to the merger that created the field elliptical
NGC~1490. Although unusual, these regions seem to fit into the generally
accepted picture regarding \HI\ in early-type galaxies. We discuss these \HII\
regions in section 2.  However, we find some evidence that some early-type
galaxies do not fit into the scheme of continued accretion. This we briefly
discuss in section 3.

\section{Intergalactic \HII\ regions near NGC 1490}

The HIPASS spectrum of NGC~1490 shows a strong detection with the presence of
almost $10^{10}$ M$_\odot$ of neutral hydrogen. This detection was followed up
with deep \HI\ observations performed with the ATCA to image the \HI\ in the
region around NGC 1490.  Five objects were detected with a total \HI\ mass of
$8.9 \times 10^9 M_\odot$ (Figure 1).  Only one \HI\ cloud has an obvious
counterpart (cloud 4) on the Digital Sky Survey in the form of a small galaxy.
The other \HI\ clouds seem to lie along an arc of more than 500 kpc long that
bends around NGC~1490.  About half of the total amount of \HI\ is found in one
cloud of about 100 kpc in size (cloud 1).  Fig.\ 2 shows a position-velocity
plot of two \HI\ clouds, as marked in Fig.\ 1.  The kinematics of the \HI\
clouds indicates that they are separate entities. E.g.\ the velocity gradient
in cloud 3 is perpendicular to the line connecting the neighbouring clouds.
The ``U-shaped'' pattern observed in the largest \HI\ cloud indicates that
this cloud is not in rotational equilibrium.

\begin{figure*}
\centerline{
\psfig{figure=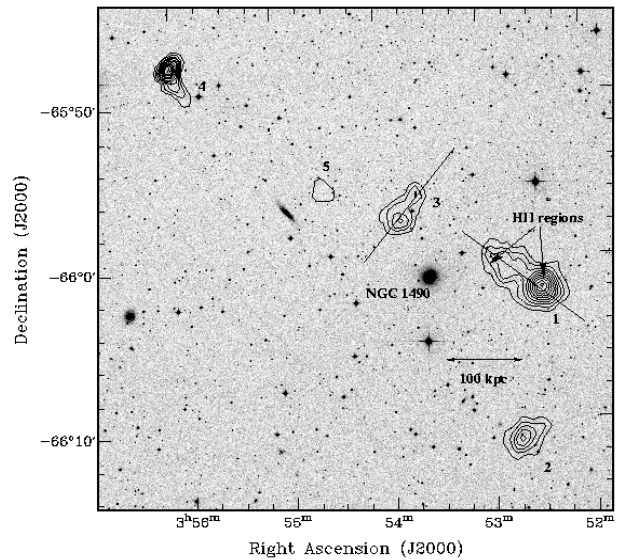,angle=-90,width=12cm}
}
\caption{View of the entire region around NGC~1490, displaying the
5 objects detected in \HI.  Contour levels: $2, 4, 6, ... \times
10^{19}$ cm$^{-2}$. The lines indicate the orientation along which the
position-velocity plots of Fig.\ 2 are taken.}
\end{figure*}

\begin{figure*}
\centerline{
\psfig{figure=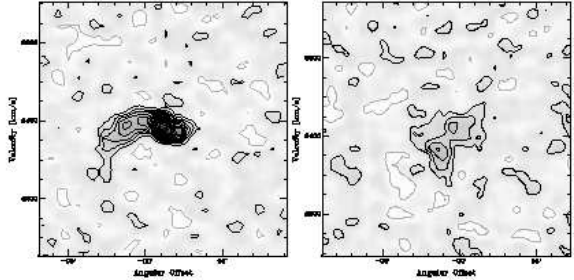,angle=-90,width=10cm}
}
\caption{ Position-velocity plots of clouds 1 and 3, taken along the line
indicated in Figure 1. Contour levels are --1, 1, 2, 3, ... mJy beam$^{-1}$}
\end{figure*}

Deep optical imaging was performed in a number of standard filters of the
region of this largest \HI\ cloud, using FORS1 on the VLT. This imaging
reveals a very low surface brightness optical counterpart (the ``OC'').  The peak
surface brightness in the V band is 24.5 V mag arcsec$^{-2}$.  The low surface
brightness nature of the OC is illustrated in Fig. 3.  The total magnitude is
about 20.5 in V.  This implies an extremely high gas-to-light ratio of $M_{\rm
HI}/L_{\rm V}
\sim 200$.  For the colour indices we find U$_{\rm B} =
0.5$, B$_{\rm V} = 0.1$ V$_{\rm R} = 0.4$ and R$_{\rm I} = 0.6$.
These colours indicate the presence of a young stellar population.

We further investigated the nature of the main \HI\ cloud and its OC
by searching for the presence of H$\alpha$ emission using narrow-band
H$\alpha$ imaging with the NTT.  These images reveal the presence of
H$\alpha$ emission in the OC at the location of the peak of the \HI,
indicating that some star formation is occurring in the OC. 
Most interestingly, two small regions of H$\alpha$ emission were found
at the edge of the main \HI\ cloud, at a distance of about 100 kpc
from the OC (indicated by 1 \& 2 in Fig. 3).  The broad-band images
indicate that there is no extended optical continuum light near these
two H$\alpha$ regions.

The H$\alpha$ luminosities of the \HII\ complexes are $3-5
\times 10^{38}$ erg s$^{-1}$ (i.e.\  30-50 times Orion).  The rate of star
formation estimated from these luminosities is very modest: around
0.005 Myr-1 per region.  All the stars corresponding to the diffuse
optical counterpart can be formed over a Hubble Time with such a star
formation rate.

Optical spectroscopy of the three H$\alpha$ regions was performed with
FORS1 on the VLT.  Preliminary analysis of the data suggests that the
abundances in the H$\alpha$ regions are only slightly below solar (in
the range 0.5-1.0 $Z_\odot$). This relatively high abundance excludes a
primordial nature of the \HI\ clouds.

\subsection{The nature of the clouds near NGC~1490}

The observations point to a tidal- or merger-related origin for the system of
\HI\ clouds near NGC 1490. The alternative hypothesis, i.e.\  the \HI\ clouds
are primordial, can be excluded because of the near solar abundance measured
in the \HII\ complexes.  The arc-like arrangement of the \HI\ clouds is
suggestive for a tidal origin.  It bears some resemblance to other systems.
There is probably a parallel between the objects discussed here and systems
like e.g.\ the Leo Ring (Schneider et al.\ 1989).  Tidal interactions between
galaxies and merging is a very diverse process that can produce very
spectacular objects with bright tidal tails.  But there are also cases where
the evidence of an interaction is only detectable in neutral hydrogen.
The extreme version is that of a galaxy being completely destroyed by a larger
galaxy without any merging taking place.  The material of the original galaxy
is spread out over such a large volume that, in the optical, the surface
brightness is well below the detection limit and only the
\HI\ remains observable (e.g.\ Ryan-Weber et al.\ 2003, see also this
proceedings).  Perhaps the \HI\ clouds near NGC~1490 have such an
origin.

\begin{figure*}
\centerline{\psfig{figure=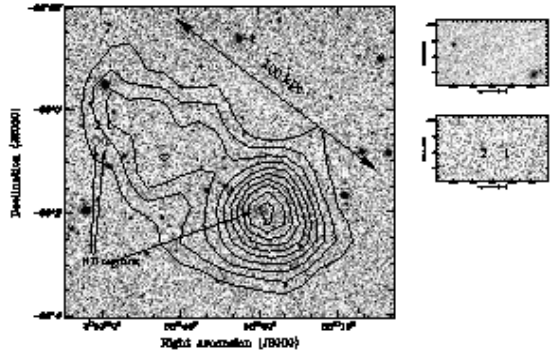,angle=-90,width=10cm}}
\caption{{\sl Left} \HI\ contours on top of an $r$-band image obtained with the
NTT. Contour levels are as in Fig. 1. The location of the three \HII\ complex
discovered using the narrow-band imaging are indicated. {\sl Right} Continuum
optical image (top) centred on the regions 1 and 2 and the continuum
subtracted line image, centred on the redshifted H$\alpha$, revealing the two
\HII\ complexes (bottom) }
\end{figure*}

An alternative hypothesis is that the \HI\ clouds are remnants of the
merger between two or more gas-rich galaxies that also created the
field elliptical NGC~1490.  The environment of NGC~1490 can be
characterised as a loose group.  The presence of an elliptical galaxy
in such a group often indicates that merging has occurred in the
group.

The interesting aspects of the case presented here are: \\ 1) the very large
amount of \HI\ involved, distributed in fairly large clouds for which only
deep optical imaging reveals an optical counterpart.  As the velocity
structure indicates, these clouds are separate entities, suggesting they are
Tidal Dwarfs even though they have no bright optical counterpart.  The clouds
are not, or hardly, visible in the optical, but, given the large reservoir of
\HI, perhaps at some point in time they will start to make stars and become
brighter optical objects. \\ 2) the existence of an extremely gas-rich object
with $M_{\rm HI}/L_V\sim 200$.  This is perhaps an example of a "Tidal Gas
Cloud" that started making stars recently.  However, it is perhaps more likely
that the optical material originates in a galaxy that was destroyed in an
interaction.
\\ 3) the presence of small \HII\ complexes that are located far away
from any optically detectable galaxy, in our case more than 100 kpc.
Only very few examples of such \HII\ regions are known.  The
observations presented here suggest that in more systems intergalactic
\HII\ complexes may exist.

\begin{figure*}
\centerline{\psfig{figure=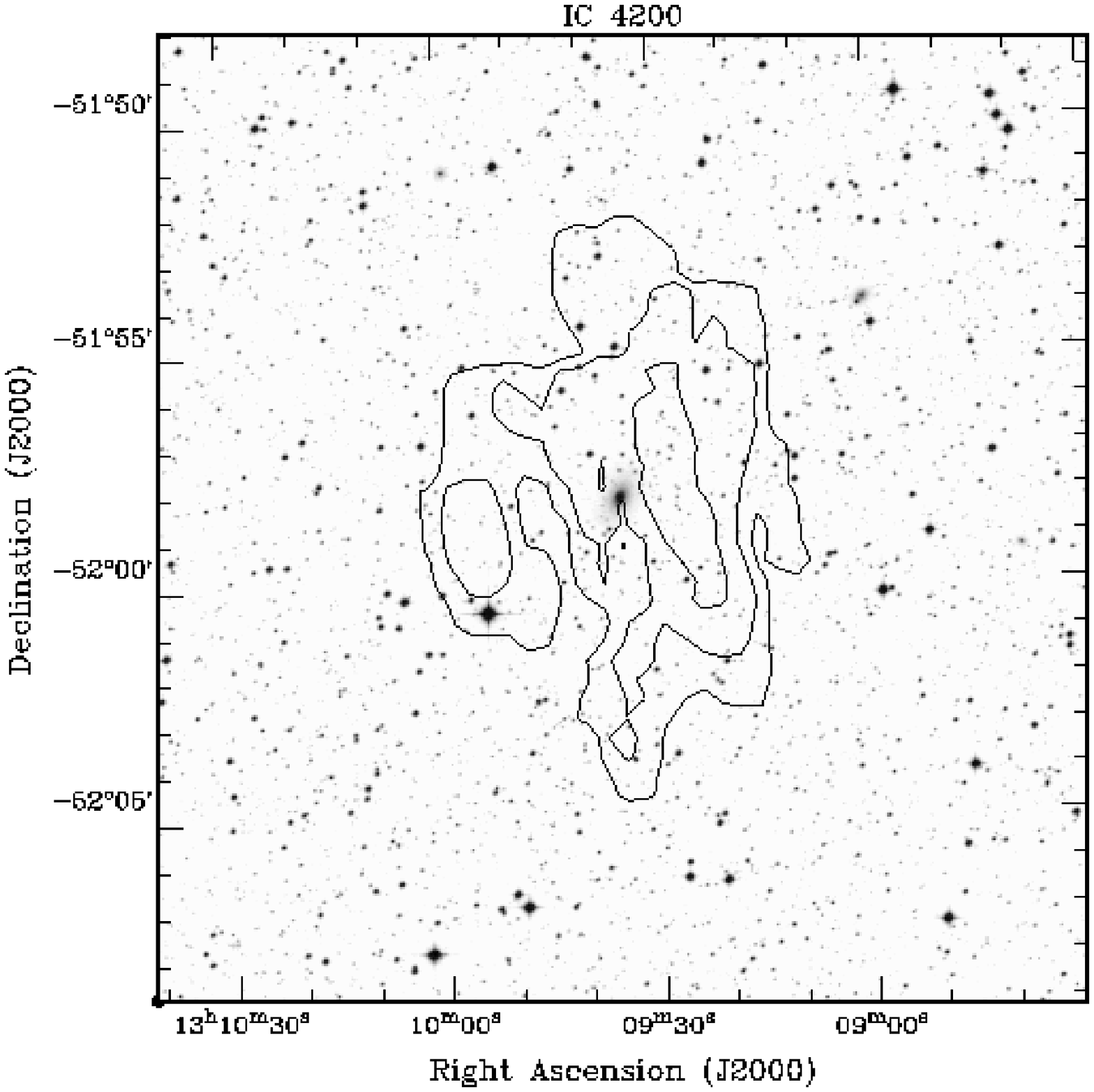,angle=-90,width=5cm}
\psfig{figure=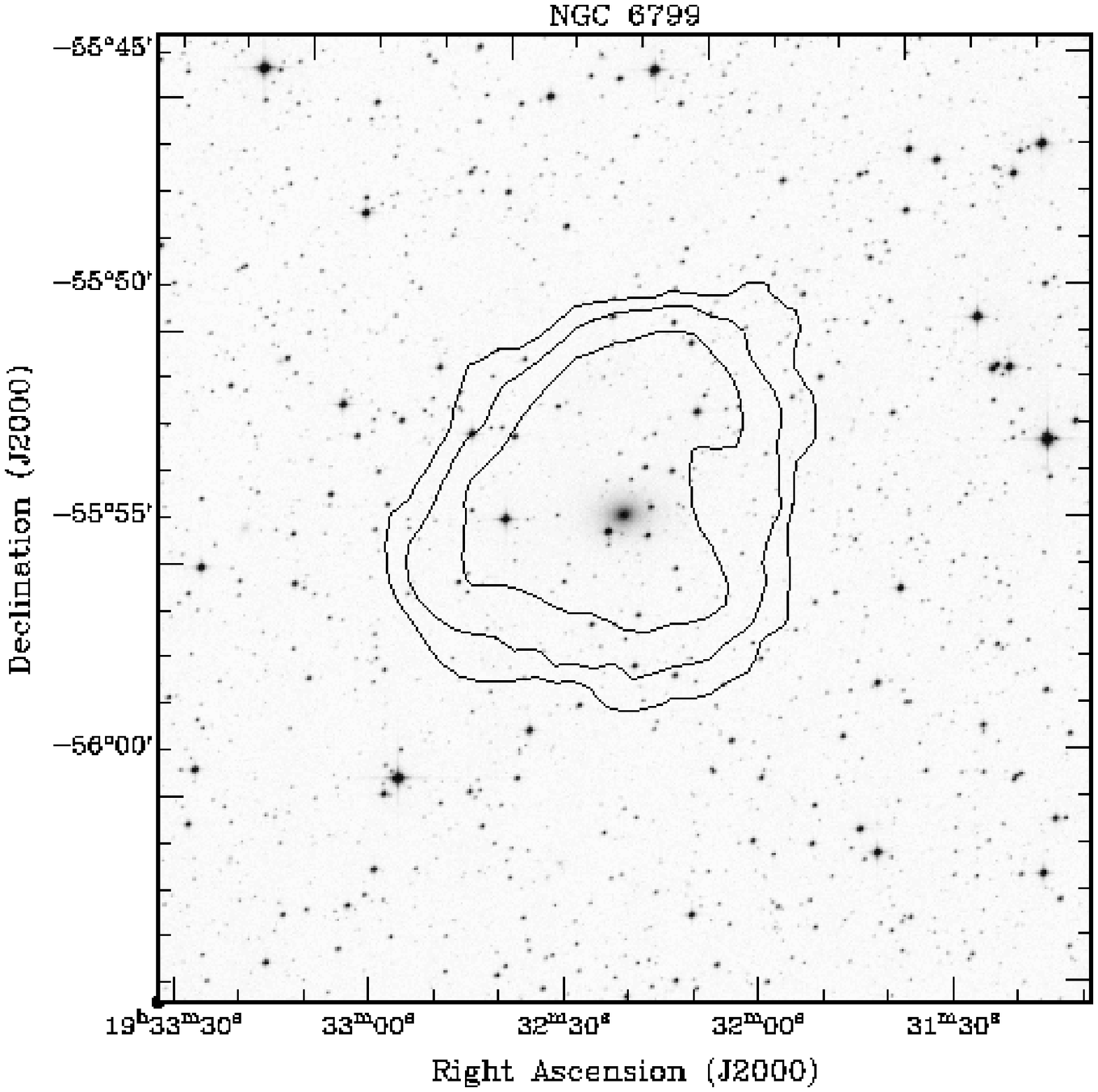,angle=-90,width=5cm}
\psfig{figure=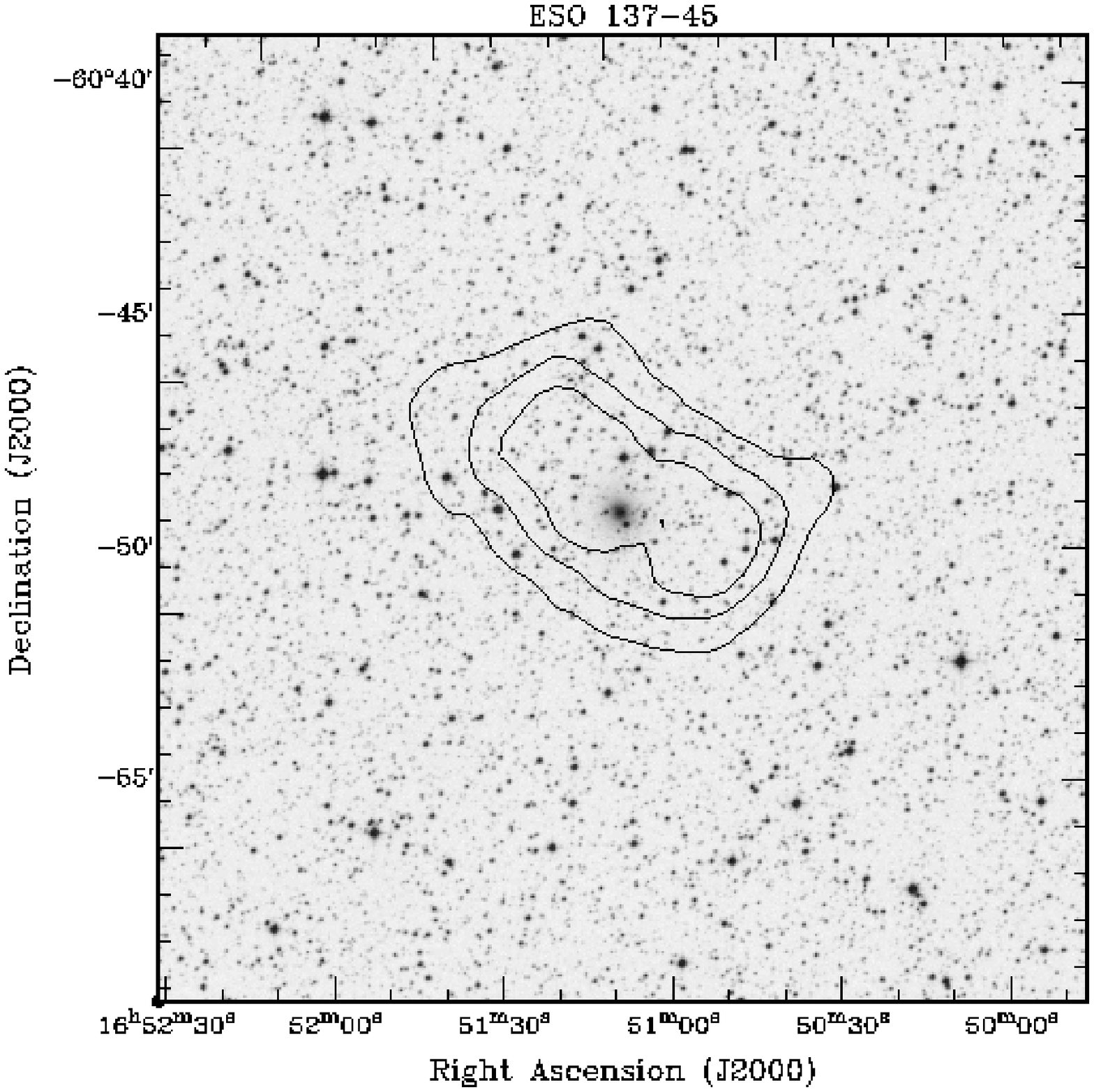,angle=-90,width=5cm}
}
\caption{Examples of very 
large regular disks of \HI\ distribution around ``normal'' early-type galaxies
obtained from the ATCA follow up of HIPASS detections (see text for details).
The disks shown here have sizes between 150 and 200 kpc. Contour levels: $2,
4, 8, 16 \times 10^{19}$ cm$^{-2}$.}
\end{figure*}

\section{Non-accretion origin of \HI\ in some early-type galaxies?}

As a final remark, we note that, although the results discussed above
fit very well into the picture continued accretion by early-type
galaxies, and the associated external origin of \HI\ in early-type
galaxies, we have found some indications that some early-type galaxies
do not fit into this scheme. Our survey has revealed a surprisingly
large number of early-type galaxies that have very large (up to 200
kpc in size), regular disks of low column density \HI.  In Fig. 4 we
present three examples of such large regularly rotating
\HI\ disks. The largest of these disks have \HI\ masses up to
10$^{10}$ M$_\odot$.  Given their size and regular appearance, these
disks must be quite old, in many cases well over $5
\times 10^9$ yr.  Hence, they are not related to recent accretions and
may be related to a much earlier formation phase of these
galaxies. Moreover, no large accretion can have occurred by these
galaxies, as these would likely have destroyed these large \HI\ disks, or have
triggered significant star formation in the \HI\ disk.
The \HI\ column density in these disks peaks at only 10$^{20}$ atoms
cm$^{-2}$, so, despite the large \HI\ reservoir, no significant star
formation is occurring. These \HI\ disks will evolve only very slowly
and will remain as purely gaseous disks for very long periods of time,
provided the environment is not too hostile for these disks.


\begin{references}

\reference Barnes et al. 2001, MNRAS 322, 486

\reference Ryan-Weber, Webster \& Bekki, 2003, ASSL Conf.  Proc.281, p223

\reference Sadler E.M., Oosterloo T., Morganti R., 2002 in {\sl The Dynamics,
Structure \& History of Galaxies},  Eds.  G.S. Da Costa and Helmut Jerjen,
ASP Conference Proceedings, Vol. 273, p.215

\reference Schneider et al.\  1989, AJ, 97, 666

\reference Schweizer F., Seitzer P., Faber S., Burnstein D., Dalle Ore C.M.,
Gonz\'ales J.J.\ 1990, ApJL 364, L33

\reference Trager S.C., Faber S., Worthey G., Gonz\'ales J.J. \ 2000, 
AJ 120, 165

\end{references}
\end{document}